\documentclass[journal]{IEEEtai}

\usepackage[colorlinks,urlcolor=blue,linkcolor=blue,citecolor=blue]{hyperref}

\usepackage{amsmath,amssymb}
\usepackage{color,array}
\usepackage{booktabs}
\usepackage{tabularx}
\usepackage{cleveref}
\usepackage{algorithm}
\usepackage{algpseudocode}

\usepackage{graphicx}

\newcolumntype{Y}{>{\raggedright\arraybackslash}X}
\newcolumntype{C}[1]{>{\centering\arraybackslash}p{#1}}
\newcommand{\ours}{\textsc{CASTLE}}

\makeatletter
\newenvironment{reprotable}{%
  \par\smallskip\noindent
  \begin{minipage}{\columnwidth}
  \def\@captype{table}%
  \centering\footnotesize
  \setlength{\tabcolsep}{3pt}%
}{%
  \end{minipage}\par\smallskip
}
\makeatother

\usepackage{cite}
\usepackage{mathtools,amssymb,amsfonts}

\usepackage[dvipsnames]{xcolor}

\begin{document}

\title{Independent Multi-Agent Reinforcement Learning with Counterfactual Semantic-Social World Models}

\author{Fernando Martinez, Tao Li\textdagger, Yingdong Lu, and Juntao Chen 
\thanks{F. Martinez and J. Chen are with the Department of Computer and Information Sciences, Fordham University, New York, NY, 10023, USA (e-mail:\texttt{\{martinezlopez,jchen504\}@fordham.edu} )}
\thanks{T. Li is with the Department of Systems Engineering, City University of Hong Kong, Hong Kong SAR, 999077, China (e-mail: \texttt{li.tao@cityu.edu.hk}) \textdagger Corresponding author.}
\thanks{Y. Lu is with IBM Research, Yorktown Heights, NY, 10598, USA (e-mail: \texttt{yingdong@us.ibm.com}).}
}

\markboth{Preprint}%
{Martinez \MakeLowercase{\textit{et al.}}: Independent MARL With Counterfactual Semantic-Social World Models}

\maketitle

\begin{abstract}
Fully decentralized multi-agent reinforcement learning (MARL), also referred to as independent learning, requires each agent to learn and act using only its local information and experience, without a centralized critic or inter-agent communication. Such a stringent information structure renders the conventional reward signal ambiguous. A poor return may result from an ineffective ego action, an incompatible teammate response, or an effective opponent response, yet scalar rewards alone do not reveal which explanation is responsible. We argue that agents can learn more effectively by prospectively comparing the consequences of candidate actions rather than diagnosing failures only from realized returns. Based on this intuition, we introduce \ours{} (Counterfactual Action-conditioned Semantic Tokens for Local Execution in Decentralized MARL), an offline-training, online-in-context guidance framework with two complementary world models. A Local Dynamics World Model, offline pre-trained over agents' local trajectories, summarizes the agent’s local trajectory dynamics and partial observability, while a Semantic-Social World Model predicts compact short-horizon task and social consequences for each candidate ego action. The latter is trained from counterfactual simulator rollouts that expose plausible teammate and opponent responses to alternative actions taken from the same logged rollout state. During online learning and execution, both world models remain frozen and are queried by agents using only locally available information. Their prediction logits provide in-context guidance to an independent PPO policy. Across 30 matched seeds on Tag, Spread, and Adversary in the benchmark multi-particle environments, our proposed CASTLE achieves the highest mean final score among the evaluated methods, exceeding the strongest baseline on each task by $10.67$, $6.46$, and $0.33$ normalized points, respectively.
\end{abstract}

\begin{IEEEImpStatement}
Agents can act independently while still depending on another to complete a task. This is relevant for autonomous teams that cannot rely on communication: each agent learns from its observations, but the effect of an action can also depend on how other agents act. This work studies whether world models trained offline can help in this setting. These models provide information on environmental dynamics and the potential short-term consequences of each available action. In simulated tasks involving cooperation and competition, our method outperformed the decentralized baselines evaluated without sharing observations or weight updates during online learning. These results indicate that prior experience can promote independent learning and may be useful for autonomous teams that cannot rely on or utilize communication.
\end{IEEEImpStatement}


\section{Introduction}
\label{sec:intro}
Multi-agent reinforcement learning (MARL) is a general learning paradigm for cooperative and competitive decision making. A key and natural aspect of MARL is that each agent acquires non-stationary experience as other agents learn to improve their policies concurrently, posing a challenge for the design and analysis of learning algorithms. Centralized training with decentralized execution (CTDE) mitigates this challenge by allowing a centralized critic or model to access the global state and joint actions during learning~\cite{amato2024introduction}. An alternative set of methods focuses on learning to communicate with another by exchanging observations, latent states, or optimization variables \cite{zhu2024survey-fd7,pan2026distributed-5c0,pan2026hessianfree}. Both of these strategies have proven successful, but neither is valid if agents must learn solely from their private local experience and then act entirely on their own. The corresponding learning paradigm is referred to as fully decentralized MARL or independent learning \cite{su2024a}. 

Since fully decentralized MARL enforces more challenging information structures \cite{tao22info}, the study of agents' independent learning dynamics, both empirically, such as Independent PPO (IPPO) \cite{de2020independent} and theoretically, such as independent policy gradient \cite{daska20idpg}, has been a recurring theme in MARL. Recent explorations include adding value transformations, decentralized policy regularization, or centralized context obtained from local observation histories~\cite{jiang2022i2q, su2024dpo, li2025rac}. However, a local trajectory does not distinguish between world dynamics and the unknown policies of other agents. An action may fail because it was poor, because a teammate chose a conflicting action, or because an opponent was coordinated. Learning these differences solely from returns can be slow and brittle. 

Since the reward signal is retrospective and tied to a single realized consequence, the independent learning agents cannot diagnose their past failures and improve future learning. Inspired by recent developments in world models \cite{hafner2025dreamerv3}, this work proposes equipping each independent agent with prospective, action-conditioned predictions from world models that incorporate latent predictive structures that compress history and predict the consequences of potential actions before they occur. Our intuition is that comparing predicted futures across different ego actions provides the agent with counterfactual evidence beyond what scalar reward signals can provide. 

Compared with existing multi-agent world models that commonly predict a centralized global state, aggregate local models centrally, or exchange predictions through a shared information cache~\cite{egorovmamba,zeng2025efficient,venugopal2024mabl,xue2026dmawm}, our proposed method achieves fully decentralized operations for both agent policy execution and world model prediction. Hence, our work applies to cooperative, competitive, and mixed-cooperative-competitive scenarios.  

Specifically, we introduce \ours{} (Counterfactual Action-conditioned Semantic Tokens for Local Execution in Decentralized MARL), an offline-training, online-guidance framework for fully decentralized MARL. In offline, each agent learns two complementary world models from offline datasets. A Local Dynamics World Model (LDWM) maps the agent's local history to a predictive representation of its own trajectory, capturing local dynamics and partial observability. A Semantic-Social World Model (SSWM) takes the local history, along with each candidate ego action, and outputs a compact token representing that action's predicted short-horizon task and social consequences. The SSWM learns these predictions from counterfactual outcomes generated by applying alternative ego actions from the same logged rollout state. Because these alternatives share a common context, they reveal how changes in ego action affect its future interactions with teammates and opponents, rather than requiring the agent to infer such differences from unrelated scalar returns. During the online stage, both world models are frozen and queried using only locally available information: the LDWM supplies predictive context to the policy, while the SSWM provides action-specific guidance to the agent's policy learning.

\begin{figure*}[t]
\centering
\includegraphics[width=0.95\textwidth]{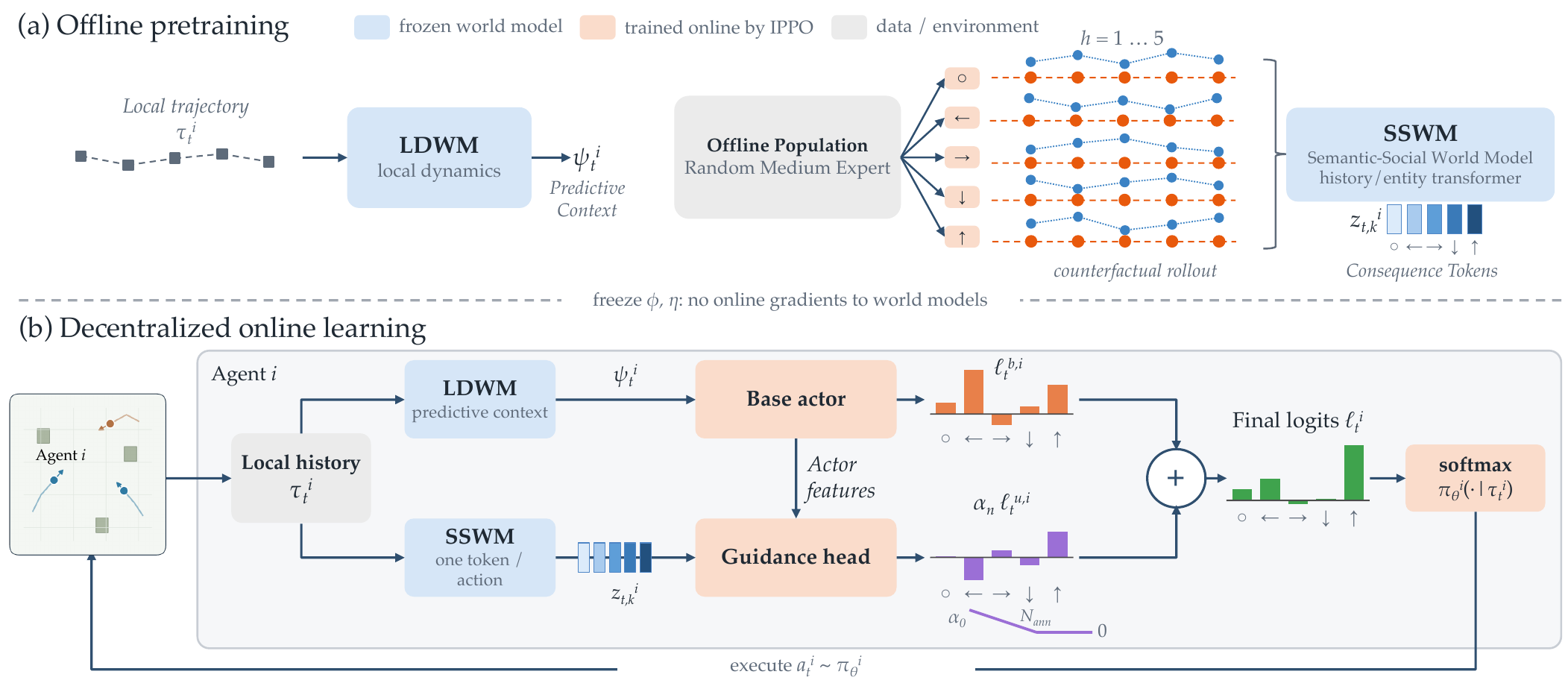}
\caption{Architecture of \ours{}. Offline pretraining produces frozen LDWM checkpoints and an SSWM that emits one consequence token $z_{t,k}^i$ for each candidate ego action. During decentralized IPPO, a trainable actor combines base policy logits with an annealed gated utility branch computed from the frozen SSWM tokens. Online gradients update only the orange PPO components; blue world-model modules remain frozen.}
\label{fig:method-overview}
\end{figure*}

Our contributions are as follows:
\begin{itemize}
    \item We introduce CASTLE, a \textbf{dual-world-model} framework for fully decentralized MARL. A Local Dynamics World Model (LDWM) summarizes dynamics and partial observability, while an action-conditioned SSWM learns counterfactual tokens that encode social consequences.
    \item We develop an offline-training, online-guidance procedure that incorporates frozen LDWM context and transient SSWM consequence guidance into conventional independent MARL. 
    \item We evaluate \ours{} using IPPO over 30 matched seeds on three cooperative and competitive MPE tasks under both in-domain and held-out pretraining. \ours{} surpasses the strongest fully decentralized baseline on Tag, Spread, and Adversary by $10.67$, $6.46$, and $0.33$, respectively. 
\end{itemize}
The rest of the paper is organized as follows. \Cref{preliminary} introduces MARL basics and notations. \Cref{sec:method} presents the proposed CASTLE framework, followed experimental evaluation in \Cref{sec:experiments}. In \Cref{sec:related}, we position our CASTLE framework within the literature on decentralized MARL (multi-agent) world models and in-context RL. Finally, \ref{sec:conclusion} concludes the paper.

\section{Preliminary}
\label{preliminary}
We model the multi-agent environment as a partially observable Markov game \cite{tao22confluence}:
\begin{equation*}
    \begin{aligned}
        \mathcal{G}=\langle
        \mathcal{N},\mathcal{S},\{\mathcal{O}^i,\mathcal{A}^i\}_{i\in\mathcal{N}},T, \{P^i,r^i\}_{i\in\mathcal{N}},\rho,\gamma
        \rangle,
    \end{aligned}
\end{equation*}
where $\mathcal{N}=\{1,\ldots,n\}$ is the set of agents. $\mathcal{S}$ denotes the state space and $\mathcal{O}^i$ and $\mathcal{A}^i$ denote the observation space and action space of agent $i$ respectively. We assume that each $\mathcal{A}^i$ is finite and discrete. Given joint action $a_t=(a_t^1,\ldots,a_t^n)$, $T( s_{t+1} \mid s_t,a_t)$ defines the state transition, $P^i(o_t^i \mid s_t)$ defines the local observation kernel, and $r^i( s_t,a_t)$ is the reward for agent $i$. The initial-state distribution is denoted by $\rho$, and $\gamma\in[0,1)$ is the discount factor. Agent $i$ selects actions based on its local history $\tau_t^i=(o_0^i,a_0^i,r_0^i,\ldots,a_{t-1}^i,r_{t-1}^i,o_t^i)$ using policy $\pi_{\theta_i}^i(a_t^i \mid \tau_t^i)$. $\tau_t=(\tau_t^1,\ldots,\tau_t^n)$ and $\theta=(\theta_1,\ldots,\theta_n)$ denote the tuple for all local histories collected and policy parameters respectively. For compactness, let $o_t=(o_t^1,\ldots,o_t^n)$ and define the joint observation kernel as $P(o_t\mid s_t)=\prod_{i\in\mathcal{N}}P^i(o_t^i\mid s_t)$. Because each agent samples its action from its own local policy, the induced joint action distribution and global trajectory distribution are
\begin{align*}
    p_{\theta}(a_t\mid\tau_t)
    &=
    \prod_{i\in\mathcal{N}}
    \pi_{\theta_i}^i(a_t^i\mid\tau_t^i),\\
    q_{\theta}(\xi)
    &=
    \rho(s_0)
    \prod_{t=0}^{\infty}
    \left[
        P(o_t\mid s_t)
        p_{\theta}(a_t\mid\tau_t)
        T(s_{t+1}\mid s_t,a_t)
    \right].
\end{align*}
where $\xi=(s_0,o_0,a_0,s_1,o_1,a_1,\ldots)$ denotes the global trajectory. Although $q_\theta$ depends on the policies of all agents, agent $i$ controls only its own parameters $\theta_i$ and seeks to maximize $J^i(\theta)=\mathbb{E}_{\xi\sim q_\theta}[\sum_{t=0}^{\infty}\gamma^t r_t^i]$ using only its local trajectory. The global trajectory $\xi$ and its distribution $q_\theta$ are analytical objects and are not available to the agents. During online learning and execution, each agent receives neither the global state, other agents' private histories, inter-agent messages, nor gradients or parameters from others.

\textbf{Independent PPO.} Our online optimizer relies on independent PPO (IPPO) \cite{schulman2017ppo}. For agent $i$, let
\begin{equation*}
    w_t^i(\theta_i)
    =
    \frac{\pi_{\theta_i}^i(a_t^i\mid\tau_t^i)}
    {\pi_{\theta_i^{\mathrm{old}}}^i(a_t^i\mid\tau_t^i)}
\end{equation*}
be the policy ratio and $\widehat{A}_t^i$ the advantage estimated from its local rollout~\cite{schulman2016gae}. The clipped actor objective is
\begin{equation}
    \begin{aligned}
        \bar{w}_t^i
        &= \operatorname{clip}(w_t^i,1-\epsilon,1+\epsilon), \\
        \mathcal{L}_{\mathrm{clip}}^i(\theta_i)
        &= \mathbb{E}_t\!\left[
            \min\!\left(
                w_t^i\widehat{A}_t^i,
                \bar{w}_t^i\widehat{A}_t^i
            \right)
        \right].
    \end{aligned}
    \label{eq:ippo-objective}
\end{equation}
where $\epsilon>0$ is the PPO clipping radius.

Agents individually calculate their own advantage and bootstrapped return target using local rewards and critic $V_{\omega_i}(\tau_t^i)$. They maximize Eq.~\eqref{eq:ippo-objective} with entropy regularization, and fit their critic to their local return target; actor and critic parameters $(\theta_i,\omega_i)$ are learned independently for each agent. \ours{} maintains this IPPO update and information structure, but alters only how the policy logits are formed: the trainable actor receives frozen predictive context with an annealed consequence correction, while both pretrained world models remain frozen.

\textbf{Recurrent state-space world models.} An RSSM represents a partially observed trajectory using a deterministic recurrent state $x_t$ and a stochastic latent state $\zeta_t$. Given the previous latent state and action, the recurrent dynamics first update the deterministic state. A dynamics predictor estimates the current stochastic state without observing $o_t$, whereas a second distribution also uses the encoded current observation~\cite{hafner2019planet}: 

\begin{equation*}
\begin{aligned}
x_t &= f_\phi(x_{t-1},\zeta_{t-1},a_{t-1}),
& p_\phi(\zeta_t\mid x_t),\\
e_t &= e_\phi(o_t),
& q_\phi(\zeta_t\mid x_t,e_t).
\end{aligned}
\end{equation*}
Thus, $q_\phi$ does not update the recurrent state ($x_t$); it estimates the current stochastic state using both the recurrent state and current observation. Prediction heads applied to the joint latent state $(x_t,\zeta_t)$ reconstruct observations and predict rewards and continuation. Algorithms such as Dreamer use the learned prior to generate imagined latent trajectories for control~\cite{hafner2025dreamerv3}. In our work, the pretrained RSSM remains frozen and its observation-conditioned latent state supplies local predictive context during online learning.

\section{Methodology}
\label{sec:method}
\noindent\textbf{Overview.} Our \ours{} separates offline world-model pretraining from online decentralized control. \Cref{fig:method-overview} provides a schematic overview of the CASTLE framework. From local trajectories, the Local Dynamics World Model (LDWM) learns a predictive representation of each agent's local dynamics. Meanwhile, the Semantic-Social World Model (SSWM) learns to predict the short-horizon consequences of every action available to the agent, given plausible behavior from its teammates and opponents. Both world models are frozen before online learning. The actor uses the LDWM context to form its base action logits and converts the SSWM consequence tokens into a temporary correction to those logits. Each agent then updates only its own actor, guidance branch, and critic through IPPO, without global state, a centralized critic, or communication.

\subsection{Counterfactual Semantic-Social World Model}

The Semantic-Social World Model (SSWM) is one of the two central components of \ours{}. Let $\tau_t^i$ denote agent $i$'s local history, which is then encoded by a transformer. For each available action $k\in\mathcal{A}^i$, an embedding of $k$ conditions a latent state that is advanced for $H$ steps using the same learned transition. Each updated latent state predicts task progress, distances to relevant entities, interaction events, and accumulated return at that horizon. These predictions concern both the agent's task and its interactions with teammates (cooperative) and opponents (competitive). A projection of the final latent state produces the consequence token $z_{t,k}^i=F_\eta(\tau_t^i,k)$. The parameters $\eta$ are learned offline and remain fixed during online learning and execution.

We construct counterfactual training targets by restarting the simulator from logged states and evaluating every available ego action from the same starting point. We refer to each rollout initialized from a restored logged state as a \emph{counterfactual simulator rollout}. For each candidate ego action $k$, we sample a behavior-policy profile from the offline population. At the first step, agent $i$ executes $k$, while the other agents sample actions from their respective behavior policies using their local information. All agents then follow the sampled behavior-policy profile for the remaining rollout steps, allowing teammate and opponent actions to adapt to the state changes induced by $k$. We run $M$ such closed-loop counterfactual rollouts per candidate action and average their outcomes at each horizon to obtain the targets:
\begin{equation}
    y_{t,k}^{i,h}
    =
    \left(
        \Delta_{t,k}^{i,h},
        d_{t,k}^{i,h},
        c_{t,k}^{i,h},
        R_{t,k}^{i,h}
    \right),
    \qquad h=1,\ldots,H.
    \label{eq:sswm-targets}
\end{equation}
Here, $\Delta$ measures role-dependent progress in the distance-based task objective; $d$ records the minimum task-relevant distance observed up to horizon $h$; $c\in[0,1]^3$ collects event indicators and smooth task-relevant interaction signals; and $R$ is the discounted return over the first $h$ steps. Let $\mathcal{U}=\{\Delta,d,c,R\}$ denote the target components, let $\mathbf{y}_t^i=\{y_{t,k}^{i,h}:k\in\mathcal{A}^i,\,h=1,\ldots,H\}$, and let $\mathcal{D}_{\mathrm{cf}}$ denote the offline dataset of pairs $(\tau_t^i,\mathbf{y}_t^i)$. We supervise these quantities at intermediate horizons and at the final horizon using the following multi-task objective:
\begin{equation}
    \mathcal{L}_{\mathrm{SSWM}}(\eta)
    =
    \mathcal{L}_{\mathrm{sem}}
    + \lambda_a \mathcal{L}_{\mathrm{act}}
    + \lambda_{\mathrm{rep}}\mathcal{R}_{\mathrm{rep}}
    \label{eq:sswm-loss}
\end{equation}
The semantic prediction loss is
\begin{equation*}
\mathcal{L}_{\mathrm{sem}}(\eta)
={}\mathbb{E}_{\mathcal{D}_{\mathrm{cf}},\,k\sim\operatorname{Unif}(\mathcal{A}^i)}
\!\left[\sum_{h=1}^{H}\sum_{u\in\mathcal{U}}\lambda_{u,h}\ell_u\!\left(\widehat{u}_{\eta,t,k}^{i,h},u_{t,k}^{i,h}\right)\right].
\end{equation*}

Here, $\widehat{\cdot}$ denotes model predictions. The expectation averages over the counterfactual dataset and a uniformly sampled candidate action, which is equivalent to averaging over all $k\in\mathcal{A}^i$. In $\mathcal{L}_{\mathrm{sem}}$, $\ell_u$ is mean squared error for the continuous targets $u\in\{\Delta,d,R\}$ and binary cross-entropy for the event vector $c$; the two sums supervise every target component at every rollout horizon. The candidate-independent action loss is $\mathcal{L}_{\mathrm{act}}=\operatorname{CE}_{\mathrm{mask}}(\widehat{p}_{-i,t},a_{-i,t})$. It predicts the logged actions of visible non-ego agents from the local-history representation and, because it does not depend on candidate action $k$, is not part of the target in Eq.~\eqref{eq:sswm-targets}. The fixed weights $\lambda_{u,h}$ also account for the additional supervision of final-horizon progress, distance, and events. We use the representation regularizer $\mathcal{R}_{\mathrm{rep}}=\max(\delta-s_z,0)^2-\lambda_{\mathrm{ent}}\mathcal{H}_{\mathrm{int}}$, where $s_z$ is the standard deviation over all candidate actions and is averaged over all minibatch samples and token dimensions. The first term penalizes variation below the threshold $\delta$. We compute $\mathcal{H}_{\mathrm{int}}$ by averaging the auxiliary interaction head's category probabilities over the minibatch and non-ego agent slots, then taking the entropy of this distribution. Weighted by $\lambda_{\mathrm{ent}}$, this term discourages concentration of the average probabilities in a single category. The auxiliary head contributes only to offline pretraining.

\subsection{Local Dynamics World Model}
In addition to the SSWM, each agent runs a frozen local copy of the Local Dynamics World Model (LDWM) $\operatorname{LDWM}_{\phi}$. The LDWM specializes the RSSM in Section~\ref{preliminary} to agent $i$'s local observation--action stream. Its deterministic state, prior, and posterior are
\begin{equation*}
\begin{aligned}
x_t^i &= f_\phi(x_{t-1}^i,\zeta_{t-1}^i,a_{t-1}^i),
&p_t^i &= p_\phi(\zeta_t^i\mid x_t^i),\\
e_t^i &= e_\phi(o_t^i),
&q_t^i &= q_\phi(\zeta_t^i\mid x_t^i,e_t^i).
\end{aligned}
\end{equation*}
The posterior distribution $\zeta_t^i\sim q_t^i$ incorporates the current local observation, and the LDWM context supplied to the actor is $\psi_t^i=[x_t^i;\zeta_t^i]=\operatorname{LDWM}_{\phi}(\tau_t^i)$. Heads conditioned on $\psi_t^i$ reconstruct $o_t^i$ and predict the local reward $r_t^i$ and continuation indicator $c_t^i$. Let $\mathcal{D}_{\mathrm{loc}}$ denote the offline dataset of agents' local trajectories and let $\bar{\cdot}$ indicate a stop-gradient. We pretrain the LDWM with the Dreamer-style objective
\begin{equation}
\begin{aligned}
\mathcal{L}_{\mathrm{LDWM}}(\phi)
={}&\mathbb{E}_{\tau^i\sim\mathcal{D}_{\mathrm{loc}}}\sum_t\bigg[
\lambda_o\mathcal{L}_{\mathrm{obs}}^t
+\lambda_r\mathcal{L}_{\mathrm{rew}}^t
+\lambda_c\mathcal{L}_{\mathrm{cont}}^t \\
&\quad+\lambda_{\mathrm{dyn}}D_{\mathrm{KL}}\!\left(\bar{q}_t^i\,\|\,p_t^i\right)
+\lambda_{\mathrm{rep}}D_{\mathrm{KL}}\!\left(q_t^i\,\|\,\bar{p}_t^i\right)
\bigg],
\end{aligned}
\label{eq:LDWM-loss}
\end{equation}
The observation and reward losses use mean squared error (MSE), and continuation prediction uses binary cross-entropy (BCE). For the reward head we use the target $\widetilde{r}_t^i=\operatorname{sgn}(r_t^i)\log(1+|r_t^i|)$ to compress reward magnitudes while keeping their signs. The three prediction objectives are
\begin{equation*}
\mathcal{L}_{\mathrm{obs}}^t
=\operatorname{MSE}(\widehat{o}_t^i,o_t^i),\;
\mathcal{L}_{\mathrm{rew}}^t
=(\widehat{r}_t^i-\widetilde{r}_t^i)^2,\;
\mathcal{L}_{\mathrm{cont}}^t
=\operatorname{BCE}(\widehat{c}_t^i,c_t^i).
\end{equation*}
In these objectives, the quantities with $\widehat{\cdot}$ are outputs of the prediction heads conditioned on $\psi_t^i$, with the reward head predicting $\widetilde{r}_t^i$. The continuation target $c_t^i$ equals one if the episode continues after step $t$ and zero otherwise. The first KL term trains the prior using $\bar{q}_t^i$ as its target, while the second trains the posterior using $\bar{p}_t^i$, balancing dynamics and representation learning~\cite{hafner2025dreamerv3,fernando26score}. During online interaction, each agent updates $(x_t^i,\zeta_t^i)$ from only its own observations and actions and passes $\psi_t^i$ to the actor encoder. The parameters $\phi$ remain fixed, and LDWM inference uses no global state, centralized critic, communication, or other agents' private information. 

\subsection{Decentralized Agent Consequence Guidance}
During online learning, the base actor learns action preferences from local experience, while the guidance branch learns how the predicted consequence of each candidate should temporarily shift those preferences. A local gate controls the overall strength of this correction. For each agent $i$, let $\theta_i$ collect the trainable parameters of its attention encoder, base actor, gate, and utility head. The attention encoder $T_{\theta}$ first combines the current local observation with the frozen LDWM context:
\begin{equation}
    h_t^i = T_\theta\!\left(o_t^i, \bar{\psi}_t^i\right).
    \label{eq:policy-trunk}
\end{equation}

The base actor outputs the action-logit vector $\boldsymbol\ell_t^{b,i}=\ell_{\theta_i}^b(h_t^i)\in\mathbb{R}^{|\mathcal{A}^i|}$. In parallel, the frozen SSWM is queried with the same local history $\tau_t^i$ and each candidate action $k$, producing $z_{t,k}^i=F_\eta(\tau_t^i,k)$. A trainable branch converts these tokens into a utility-logit vector:
\begin{equation}
\begin{aligned}
    g_t^i
    &= g_\sigma\!\left(g_\theta(\bar{h}_t^i)\right), &
    u_{t,k}^i
    &= u_\theta\!\left(\bar{h}_t^i,\bar{z}_{t,k}^i,e_k\right), \\
    \ell_{t, k}^{u,i}
    &= g_t^i u_{t,k}^i, &
    \boldsymbol\ell_t^{u,i}
    &= [\ell_{t,k}^{u,i}]_{k=1}^{|\mathcal A^i|}, \\
    \boldsymbol\ell_t^i
    &= \boldsymbol\ell_t^{b,i}
    + \alpha_n\boldsymbol\ell_t^{u,i}.
\end{aligned}
\label{eq:final-logits}
\end{equation}
Here, $g_t^i$ is a scalar gate computed from the agent's local actor representation, $g_\sigma$ denotes a scaled logistic sigmoid (see Table~\ref{tab:castle-architecture} for details),  $u_{t,k}^i$ is an action-specific utility score, and $e_k$ is a one-hot encoding of candidate action $k$. The SSWM token $z_{t,k}^i$ is a vector-valued representation of predicted consequences of taking action $k$, while a categorical policy requires one scalar preference per action. The candidate utility branch thus maps the token and local actor representation to a scalar correction to apply in the actor's logit space. The stop-gradient on $h_t^i$ prevents the guidance branch from reshaping the shared actor representation through this path; the base actor continues to train that representation through PPO. The gate and utility head are learned solely by backpropagating the IPPO actor objective through the final policy. They have no auxiliary loss providing target values for $u_{t,k}^i$. In this way, the frozen SSWM provides predictive features while online experiences determine how those features influence action preferences. 

Let $\pi_b^i(\cdot\mid\tau_t^i)=\operatorname{softmax}(\boldsymbol\ell_t^{b,i})$ be the base IPPO policy conditioned on the LDWM context. After normalization over the available actions, the policy after the $n^{th}$ PPO update satisfies
\vskip -0.5cm
\begin{align}
\pi_\theta^i(k\mid\tau_t^i)
\propto
\pi_b^i(k\mid\tau_t^i)
\exp\!\left(\alpha_n\ell_{t,k}^{u,i}\right),
\qquad k\in\mathcal{A}^i.
\label{eq:guided-policy}
\end{align}
Consequence guidance, therefore, updates the distribution over actions with learned utility-logit vector from online experience, instead of learning a separate policy. When $\alpha_n=0$, this additive correction vanishes, and Eq.~\eqref{eq:guided-policy} reduces to the base actor policy after normalization.

The utility logit $\ell_{t,k}^{u,i}$ is the action-specific score obtained by mapping the SSWM
token through the learned utility branch. Because categorical logits represent
relative action preferences, adding $\boldsymbol\ell_t^{u,i}$ makes it a residual
correction to the base actor rather than a second independently normalized
policy. In particular, Eq.~\eqref{eq:guided-policy} gives
\begin{equation*}
\log\frac{\pi_\theta^i(k\mid\tau_t^i)}
         {\pi_\theta^i(j\mid\tau_t^i)}
=
\log\frac{\pi_b^i(k\mid\tau_t^i)}
         {\pi_b^i(j\mid\tau_t^i)}
+\alpha_n\left(\ell_{t,k}^{u,i}-\ell_{t,j}^{u,i}\right).
\end{equation*}
The SSWM branch therefore reranks the base actor's candidate actions according
to differences in their predicted consequence scores, while preserving the
base preferences when those scores are equal.

For $\alpha_n>0$, this logit addition is equivalently the solution of
\begin{equation*}
\begin{aligned}
\pi_\theta^i(\cdot\mid\tau_t^i)
={}&\underset{q\in\Delta(\mathcal A^i)}{\arg\max}\;
\left\{\alpha_n\sum_{k\in\mathcal A^i}q(k)\ell_{t,k}^{u,i} -D_{\mathrm{KL}}\!\left(q\,\|\,\pi_b^i(\cdot\mid\tau_t^i)\right)
\right\}.
\end{aligned}
\end{equation*}
Thus, the correction has the geometry of a KL-regularized policy-improvement step: it favors actions with higher consequence scores while penalizing departure from the locally informed base policy. To see when this interpretation corresponds to genuine improvement, suppose that
$\ell_{t,k}^{u,i}=A^{\pi_b^i}(\tau_t^i,k)$ is the exact advantage of the base policy. Since the base policy is feasible and has zero expected advantage, optimality of the guided policy implies
\begin{equation*}
\mathbb E_{k\sim\pi_\theta^i}\!\left[
A^{\pi_b^i}(\tau_t^i,k)\right]
\geq \frac{1}{\alpha_n}
D_{\mathrm{KL}}\!\left(\pi_\theta^i\,\|\,\pi_b^i\right)
\geq 0.
\end{equation*}
Holding the other agents' policies fixed, an exact advantage correction would therefore yield a local policy-improvement step. In our method, however, $\ell_{t,k}^{u,i}$ is not supervised to equal this advantage. Instead, IPPO learns from online returns how the frozen counterfactual features should affect action preferences. Consequently, logit addition supplies the conservative update geometry, while improvement depends on whether the learned correction becomes advantage-aligned. The context gate and the annealing coefficient limit the effect of inaccurate corrections, and $\alpha_n\to0$ returns the policy to $\pi_b^i$.

The guiding signal coming from the utility head is gated by both the local-context gate $g_t^i$ and $\alpha_n$, instead of having heterogeneous semantic predictions directly added to the actor logits. We linearly anneal $\alpha_n$ from $\alpha_0$ to zero over $N_{\mathrm{ann}}$ PPO updates. This way, the SSWM heavily influences action selection at the beginning of training, but its direct impact on the policy logits disappears once annealed. Both world models continue to run during the forward pass, and the LDWM continues to inject predictive context into the base actor. For each agent, we jointly optimize actor, guidance branch, and local critic with IPPO using only each agent's rollout, while keeping both pretrained world models fixed~\cite{yu2022ppo}.

Algorithm~\ref{alg:castle} summarizes offline pretraining and each agent's independent online updates.
\begin{algorithm}[ht]
\caption{\ours{}
}
\label{alg:castle}
\small
\begin{algorithmic}[1]
\Require Offline local trajectories, logged simulator states and joint actions, behavior-policy population; $H$, $M$, $\alpha_0$, $N_{\mathrm{ann}}$.
\Statex \textbf{Offline pretraining}
\For{each logged $(s_t,i)$ and $k\in\mathcal{A}^i$}
    \For{$m=1,\ldots,M$}
        \State Sample behavior-policy profile $\beta^{(m)}$; restore $s_t$.
        \State Execute $a_t^i\gets k$ and sample $a_t^{-i}\sim\beta_{-i}^{(m)}$.
        \State Roll out the remaining $H-1$ steps with all agents following $\beta^{(m)}$.
        \State Compute $(\Delta,d,c,R)$ through each horizon $h=1,\ldots,H$.
    \EndFor
  \State $y_{t,k}^{i,h}\gets$ mean counterfactual-rollout targets at each $h$ (Eq.~\eqref{eq:sswm-targets}).
\EndFor
\State Minimize $\mathcal{L}_{\mathrm{LDWM}}(\phi)$ on local trajectories (Eq.~\eqref{eq:LDWM-loss}).
\State Minimize $\mathcal{L}_{\mathrm{SSWM}}(\eta)$ using offline targets (Eq.~\eqref{eq:sswm-loss}).
\State Freeze $\phi,\eta$.
\Statex \textbf{Online, independently for each agent $i$}
\State Initialize $\theta_i$, $\omega_i$, optimizers, local history, and recurrent state.
\For{PPO updates $n=0,1,\ldots$ within the interaction budget}
    \State $\alpha_n\gets\alpha_0\max\{0,1-n/N_{\mathrm{ann}}\}$.
    \For{each local rollout step $t$}
        \State $\psi_t^i\gets\operatorname{LDWM}_{\phi}(\tau_t^i)$.
        \State $z_{t,k}^i\gets F_\eta(\tau_t^i,k)$, $k\in\mathcal{A}^i$.
        \State $h_t^i\gets T_{\theta_i}(o_t^i,\bar{\psi}_t^i)$; compute $\ell_t^i$ using Eq.~\eqref{eq:final-logits}.
        \State $a_t^i\sim\operatorname{softmax}(\ell_t^i)$.
        \State Execute $a_t^i$; observe $r_t^i$, $o_{t+1}^i$, and termination.
        \State Store local transition and $\log\pi_{\theta_i}^i(a_t^i\mid\tau_t^i)$.
        \State Advance local history; reset history and recurrent state at episode boundaries.
    \EndFor
    \State Compute $\widehat{A}_t^i$ and local return targets using $V_{\omega_i}$.
    \For{PPO epochs and local minibatches ($\alpha_n$ fixed)}
\State Update $\theta_i$ via Eq.~\eqref{eq:ippo-objective} with entropy regularization.
        \State Fit $V_{\omega_i}$ to the local return targets.
    \EndFor
\EndFor
\end{algorithmic}
\end{algorithm}
\section{Experiments}
\label{sec:experiments}
We evaluate three questions: whether offline counterfactual pretraining improves fully decentralized online learning, whether the learned consequence representation transfers when the target environment is excluded from pretraining, and how the composition of the offline population affects performance. All results use fixed online training budgets and are reported without checkpoint selection. 

\subsection{Offline Data and Pretraining}
\begin{figure*}[ht]
\centering
\includegraphics{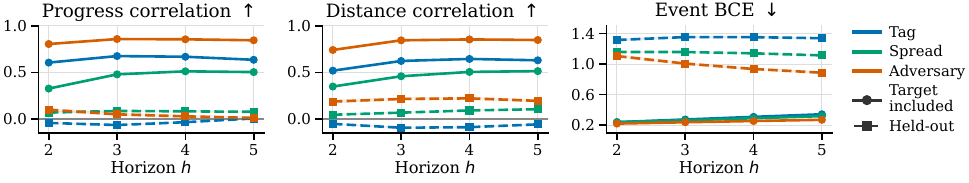}
\caption{Target-task predictive accuracy over the SSWM rollout horizon. Solid lines indicate target-included pretraining and dashed lines held-out pretraining, evaluated on identical target-task histories. Correlations are computed across candidate actions; shading denotes 95\% confidence intervals.}
\label{fig:sswm-pretraining-diagnostics}
\end{figure*}
We generate task-specific offline populations by running random policies and policy checkpoints spanning different levels of competence. Tag and Spread populations combine random, decentralized IPPO, and MAPPO sources, while Adversary populations combine random and MAPPO sources. We retain these behavior-policy checkpoints for counterfactual label generation. Recorded for each transition and acting agent are the local observation, previous and current action, reward, termination flag, visible-entity features, and policy-source metadata. The simulator state, policy-source metadata, and features used to compute $\mathbf{y}_t^i$ support only offline label construction; the SSWM itself receives only $(\tau_t^i,k)$. MAPPO checkpoints are likewise used only to generate offline trajectories and counterfactual rollout labels; centralized training information is never available to the online agents. Random policies sample actions uniformly. Medium and expert labels follow the training stage for MAPPO and recorded source-policy performance for decentralized IPPO, as summarized in Appendix Table~\ref{tab:offline-population}.

For each logged state and acting agent, we enumerate all five possible actions and use counterfactual simulator rollouts to construct the labels in Eq.~\eqref{eq:sswm-targets}. Each action is evaluated through four closed-loop counterfactual rollouts using a five-step lookahead. In each rollout, we sample a behavior-policy profile from the offline population. The acting agent executes the candidate action at the first step, while the other agents act according to the sampled profile; thereafter, every agent follows that profile using its updated local information. Results are averaged across the four rollouts. These behavior policies need not match the agents encountered online, but their diversity exposes the SSWM to plausible teammate and opponent responses to the state changes induced by each candidate ego action.

Both frozen models are pretrained separately. The LDWM is pretrained to reconstruct local observations, predict rewards and continuation, and regularize its latent dynamics using only local transitions; it is never conditioned on joint observations nor given access to the global state. The SSWM optimizer is run with Eq.~\eqref{eq:sswm-loss} on source-stratified offline batches. One deterministic subset of offline examples is held out from gradient updates and used exclusively for validation. We check a series of prespecified validation metrics before deploying each checkpoint to confirm that it preserves candidate-wise progress and distance rankings, predicts interaction events and non-ego actions, and maintains non-degenerate candidate-token variation and auxiliary interaction-head usage. Online returns are used to neither train nor select either world model.

We present the semantic prediction quality at each stage of the rollout horizon in Figure~\ref{fig:sswm-pretraining-diagnostics}. When including target-task data in pretraining, candidate-distance rankings remain informative through horizon $h=5$, although event-prediction error increases with rollout depth. Predictive ranking is strongest on Adversary, remains stable on Tag, and improves over the early horizons on Spread. When the target task is excluded from pretraining, candidate-distance correlation approaches zero on Tag and Spread, while event-prediction error increases on all three tasks. Because both target-task included and held-out checkpoints are validated on the exact same held-out target-task histories, this diagnostic isolates predictive transfer from downstream online policy performance.

\begin{table*}[t!]
\centering
\caption{D4RL-normalized final performance over 30 matched seeds, reported as mean $\pm$ standard deviation. Higher is better. In-domain (R) adds target-task random trajectories to the held-out source pool, while in-domain (R+M) adds target-task random and intermediate trajectories. Bold and underlined entries are the best and second-best means in each column.}
\label{tab:main-setting-results}
\begin{tabularx}{\textwidth}{
    l
    >{\centering\arraybackslash}X
    >{\centering\arraybackslash}X
    >{\centering\arraybackslash}X
    >{\centering\arraybackslash}X
}
\toprule
Method
& \texttt{tag}
& \texttt{spread}
& \texttt{adversary}
& Total Average \\
\midrule

\multicolumn{5}{l}{\textit{Decentralized baselines}} \\
IPPO
& $\underline{78.07} \pm 8.67$
& $70.69 \pm 12.29$
& $100.34 \pm 0.92$
& $83.03$ \\

IPPO + Transformer
& $70.83 \pm 10.56$
& $79.64 \pm 5.58$
& $102.91 \pm 1.41$
& $84.46$ \\

IQL
& $76.75 \pm 7.48$
& $76.41 \pm 9.24$
& $95.55 \pm 1.84$
& $82.90$\\

IQL + Transformer
& $65.91 \pm 8.04$
& $\underline{86.11} \pm 4.33$
& $101.05 \pm 1.33$
& $84.36$\\

DPO
& $74.46 \pm 7.31$
& $82.03 \pm 4.47$
& $98.77 \pm 1.22$
& $85.09$ \\

Dreamer
& $57.67 \pm 11.31$
& $24.26 \pm 15.15$
& $\underline{105.48} \pm 0.94$
& $62.47$ \\

\midrule
\multicolumn{5}{l}{\textbf{\ours{} (ours)}} \\

Held-out
& $72.41 \pm 11.45$
& $84.72 \pm 6.97$
& $105.19 \pm 1.12$
& $87.44$ \\

In-domain (R)
& $72.19 \pm 10.98$
& $81.48 \pm 7.10$
& $103.40 \pm 0.83$
& $85.69$ \\

In-domain (R+M)
& $\mathbf{88.74} \pm 9.19$
& $\mathbf{92.57} \pm 1.65$
& $\mathbf{105.81} \pm 0.76$
& $\mathbf{95.70}$ \\

\bottomrule
\end{tabularx}
\end{table*}

\subsection{Experiment Setup}
\label{subsec:evaluation-protocol}

We use three Multi-Agent Particle Environment (MPE) tasks~\cite{flair2024jaxmarl}: Tag, Spread, and Adversary. In Tag, three cooperative predators try to catch one faster prey agent. In Spread, three homogeneous cooperative agents cover three landmarks while avoiding colliding into each other. In Adversary, two informed cooperative agents move towards a target landmark while an uninformed adversary tries to infer which landmark is the target and move there as well. These tasks require learning pursuit behavior, cooperative allocation behavior, and deceptive partial-information cooperative behavior, respectively. 

Across all methods, we use the same decentralized information setting: agents cannot communicate with each other, learning is restricted to information from the locally available history, and no centralized critic is used. We compare against IPPO, IPPO with a Transformer encoder, IQL, IQL with a Transformer encoder based on JaxMARL implementation~\cite{flair2024jaxmarl}, DPO~\cite{su2024dpo}, and a decentralized Dreamer implementation in which each agent uses separate parameters and maintains its own replay buffer. For each target task, held-out SSWM pretraining employs random, medium, and expert trajectories drawn from the other two MPE tasks, excluding any from the task itself. The in-domain (R) method integrates only random trajectories from the target task into the source pool, contrasting with in-domain (R+M), which also leverages intermediate target-task trajectories.


\subsection{Online Performance}
\label{subsec:main-online-performance}

Table~\ref{tab:main-setting-results} shows D4RL-normalized final-window results. The \ours{} columns consist of the framework pictured in Figure~\ref{fig:method-overview}: LDWM context frozen along with annealed SSWM consequence guidance. The three columns differ by the three \ours{} variants differ in the target-task behavior available during offline pretraining, as defined in Section~\ref{subsec:evaluation-protocol}.

Our in-domain (R+M) variant reaches the highest mean across all three tasks and the highest mean normalized score across tasks. Compared to the strongest decentralized baseline in each row, it performs $10.67$, $6.46$, and $0.33$ points higher on Tag, Spread, and Adversary, respectively.

\begin{figure*}[ht]
\centering
\includegraphics[
    width=1\textwidth
]{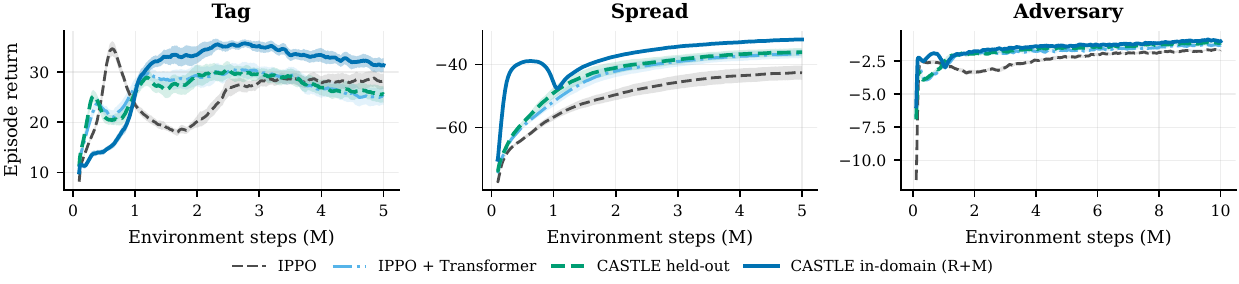}
\vspace{-15pt}
\caption{Raw online returns over 30 matched seeds. Lines show means and shaded regions 95\% confidence intervals. Curves use a 25-update moving average.}
\label{fig:main_lc}
\end{figure*}

Figure~\ref{fig:main_lc} shows that consequence guidance alters learning across regimes of interaction. During Tag, \ours{}'s in-domain learning starts slowly but enters a higher-performance region after about a million steps. During Spread, learning accelerates immediately, and the transient around one million steps results from the conclusion of the guidance schedule; then \ours{} continues fully on decentralized IPPO and maintains its lead. These curves are closer together for Adversary, indicating that guidance is honing in on the terminal policy more for this task. Held-out guidance shows a broadly similar pattern but does not always confer an advantage.

\subsection{Held-Out and In-Domain Transfer}
\label{subsec:heldout-indomain-transfer}
Target-task exposure alone is not enough to see gains in transfer. Compared to held-out pretraining, adding only random and weak target-task trajectories changes normalized score by $-0.22$ on Tag, $-3.24$ on Spread, and $-1.79$ on Adversary. Improving on this baseline by adding intermediate-expert target-task behavior on top of random data further raises scores by $16.55$, $11.09$, and $2.41$ points, respectively. The resulting in-domain (R+M) improvements over held-out pretraining are therefore $16.33$ points on Tag, $7.85$ points on Spread, and $0.62$ points on Adversary. The large Tag gap shows that pursuit particularly benefits from coverage of target-task interaction. By contrast, the small gaps on Spread and Adversary indicate that the consequence representation learned by interacting with the other MPE tasks already includes much of the structure needed to solve those tasks. 
\subsection{Ablation Studies}
\label{subsec:ablation}
\begin{figure*}[ht]
\centering
\includegraphics{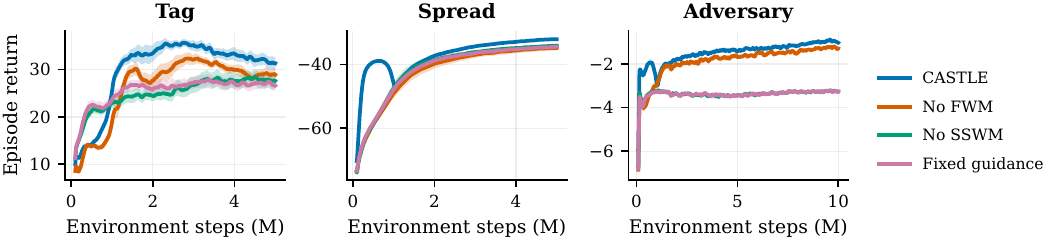}
\caption{Raw online returns over 30 matched seeds. Lines show means and shaded regions 95\% confidence intervals. Curves use a 25-update moving average.}
\label{fig:ablation_lc}
\end{figure*}

The primary comparison shows the efficacy of the entire method but does not attribute which design choices provide the gain. Here we compare CASTLE in-domain (R+M) against three controls: one that removes LDWM context, one that removes SSWM consequence guidance, or one that keeps the SSWM correction module active throughout training rather than annealing it out. All the hyperparameters are kept the same, and all applicable pretrained components are from the same in-domain (R+M) offline population. Figure~\ref{fig:ablation_lc} shows the resulting learning curves.

These curves demonstrate different benefits between interaction regimes. CASTLE improves more slowly before separating after about a million steps on Tag, thereafter having the strongest return through the rest of training. On Spread it experiences a large early improvement before producing an aligned deviation upon removal of guidance. This behavior is exhibited across all 30 seeds. Decentralized IPPO then recovers to maintain an advantage later in training. On Adversary, both CASTLE and the configuration without LDWM show continued improvement while the no-SSWM and fixed-guidance configurations stay at levels close to their early plateau.

\subsection{Mechanism Diagnostics}
\label{subsec:mechanism-diagnostics}

The ablations demonstrate whether the SSWM branch impacts learning, however they leave open the question of how the branch alters action selection behavior. To answer this question, we compare two policy distributions at every update. The guided policy $\pi_t^i$ is the distribution used to sample actions from the final logits of Eq.~\eqref{eq:final-logits}. The corresponding base policy $\pi_t^{b,i}$ is obtained from the same actor given the same local input, but after removing the SSWM utility correction. We quantify their difference with $D_{\mathrm{KL}}(\pi_t^i\,\|\,\pi_t^{b,i})$, averaged across local rollout samples. If this quantity is zero, then the branch leaves the base policy unchanged, whereas a larger value indicates a stronger redistribution of action probability.

We next measure whether the policy redistribution induced by guidance is aligned with the subsequent local learning signal. For the sampled action, we compute $\Delta\log\pi_t^i=\log\pi_t^i(a_t^i)-\log\pi_t^{b,i}(a_t^i)$ and correlate it with the local GAE estimate $\widehat{A}_t^i$ obtained from the completed rollout~\cite{schulman2016gae}. A positive correlation indicates that guidance tends to increase the probability of actions assigned higher local advantage. This correlation measures alignment with PPO's online learning signal rather than an independent causal effect; causal evidence is provided by the matched ablations in Section~\ref{subsec:ablation}. This diagnostic does not control the utility gate or update either frozen world model. Guidance is disabled after update 244, after which the policy is optimized exclusively through decentralized IPPO.

\begin{figure}[h]
    \centering
    \includegraphics[
    width=1\linewidth]{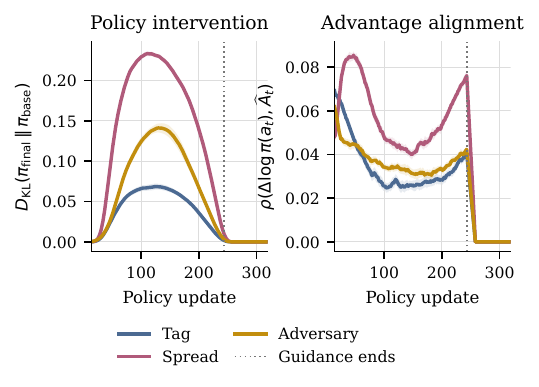}
    \vspace{-15pt}
\caption{Effect of consequence guidance over 30 matched seeds. \textbf{Top}: KL divergence between the guided policy and the base IPPO actor. \textbf{Bottom}: correlation between the guidance-induced change in the sampled action's log probability and its realized local GAE advantage.}
    \label{fig:guidance_mechanism}
\end{figure}

Despite monotonically decreasing throughout training, in Figure~\ref{fig:guidance_mechanism} policy KL first increases until updates $126$, $111$, and $130$ for Tag, Spread, and Adversary, respectively, before decreasing again. The peak KL values are $0.069$, $0.233$, and $0.141$. Intervention is therefore not solely based on the annealing schedule. Intervention also depends on how the trainable gate and utility head learn to utilize the frozen tokens given the evolving policy and visited state distribution. The subsequent drop happens when the diminishing coefficient begins to outweigh this learned scaling. Finally, the near-zero KL following update 244 exists mostly to verify that annealing actually cuts off the branch as intended.

Alignment shows a correlated result. Looking solely at positive or aggregate values, alignment stays positive throughout this period, showing mean alignment scores of $.0359$ on Tag, $.0584$ on Spread, and $.0383$ on Adversary. Spread shows both the highest intervention and alignment makes sense given its quick early gains. However, Tag achieves the highest final performance improvement compared to the decentralized baselines while having the lowest KL by far, so intervention size does not track final performance across tasks. Figure~\ref{fig:guidance_mechanism}, therefore, verifies that the branch becomes active, adapts to the online PPO signal, and is removed by annealing. Whether SSWM guidance improves learning is a separate question addressed by the matched component controls in Section~\ref{subsec:ablation}.

\subsection{Discussion}
\label{subsec:results-discussion}
The results indicate that SSWM guidance is most effective here when it shapes early action selection rather than maintaining a persistent influence throughout training. Maintaining SSWM correction performs worse than annealed guidance; once the correction weight decays to zero, subsequent policy updates rely exclusively on local experience, while the frozen LDWM continues to provide context. While the idea of distilling inductive bias into RL agents is not new \cite{silver19bias,li2025reinforcement,lei25dt-rl}, our world model approach provides in-context bias on the logits level. Although evaluated with decentralized IPPO, the framework is not tied to PPO's clipped objective and could be adapted to other decentralized policy-gradient methods with categorical policies. 

These results should be understood in the context of our experimental setup. We benchmark three specific discrete-action MPE tasks, and our current label-construction pipeline requires offline access to simulator states from which counterfactual simulator rollouts can be initialized. Its semantic targets rely on geometric relations and interaction events provided by these tasks. As both world models' weights remain unchanged during online learning, their predictions can degrade when the policy experiences local histories not frequently encountered in the offline data. Annealing limits how long the SSWM has direct effect over action selection; however, it is agnostic to whether a prediction is trustworthy. Future work can estimate predictive uncertainty from local information and scale down guidance as the prediction becomes less certain, while maintaining decentralized execution.

\section{Related Works}
\label{sec:related}
\subsection{Fully Decentralized Multi-Agent Reinforcement Learning}
CTDE methods optimize with access to privileged joint information and execute local policies after training \cite{yu2022ppo}. Fully decentralized methods train without a centralized signal. IPPO decentralizes by independently applying PPO \cite{yu2022ppo}. I2Q and Best Possible Q-Learning reformulate the local value-learning objective to lessen transitions' dependence on other agents' actions \cite{jiang2022i2q,jiang2025bql}. DPO backpropagates a decentralized surrogate signal for multi-agent policy improvement \cite{su2024dpo}. RAC conditions each local policy on return-derived context to help infer unobserved changes in joint behavior \cite{li2025rac}. The term decentralized has also been used in a separate literature to refer to networked optimization where agents can exchange gradients or parameters with neighbors \cite{zhang2018networked,chen2022communication}. Our online learning setting rules out both centralized and networked training signals, \ours{}, freezes and locally computes action-consequence tokens for each independent learner and diminishes their direct influence as online experience is collected.

\subsection{Multi-Agent World Models}
Single-agent world models learn compact latent states for planning or learning policies \cite{hafner2019planet,hafner2025dreamerv3,schrittwieser2020muzero}. Multi-agent variants additionally need to model how agents affect the transition function. MAMBA and MABL learn compositional multi-agent latent dynamics with CTDE \cite{egorovmamba,venugopal2024mabl}, MARIE performs decentralized sequence modeling with centralized aggregation \cite{zhang2025marie}. MAG instead studies agent-wise local models and explicitly optimizes how their prediction errors interact over multi-step rollouts \cite{wu2023models}. Recent work also leverages diffusion models \cite{zhang2025dima}, mixture-of-experts \cite{zhao2025m3w}, and compositional models \cite{zhang2025combo} for joint or agent-structured imagination for control and planning. Other decentralized world-model approaches rely on explicit information sharing during training \cite{zeng2025efficient}. In contrast, \ours{} does not use a learned global latent state or online messages. Its world models are pretrained offline, queried from local histories, and frozen throughout decentralized policy optimization.

Counterfactual models of other agents have been used to measure social influence as well. Jaques et al. predict how alternative actions of one agent change the action distribution of another agent and use this causal influence as an intrinsic reward \cite{jaques2019social}. \ours{} instead predicts the full range of physical, task, and social consequences of every action and lets the online actor learn how those predictions should alter its action preferences. It neither defines an influence reward nor uses online messages or a global latent state.

\subsection{Offline Pretraining and In-Context Reinforcement Learning}
Inspired by recent advances in large language models, in-context learning lets a model use examples provided in the input to condition its predictions without changing its parameters~\cite{dong-etal-2024-survey}. In RL, trajectories collected during recent experience can provide this conditioning context, allowing a pretrained policy to reuse information between trials or adapt to changes in the task and other agents. Recent methods accomplish this with hindsight-organized experience, hierarchical in-context decisions, locally consistent predictions of other agents, or obtain context from a pretrained communicative world model~\cite{liu2023emergent,huang2024decision,tao23sce,martinez2025context}. 

Offline MARL has likewise leveraged centralized-to-decentralized distillation, diffusion policies, and sequence models to extract coordinated behavior from fixed datasets~\cite{tseng2022distillation,zhu2024madiff,formanek2025oryx}. These approaches place offline knowledge directly into agents' action-generating policies, leaving coverage and quality of the pretraining dataset as primary determinants of learned behavior \cite{formanek2024mirage,tao24col,kim-tao25col}.

\section{Conclusion}
\label{sec:conclusion}
We presented \ours{}, an offline-to-online world-model based method for fully decentralized MARL. A frozen LDWM provides predictive dynamics context to each agent's actor, while the frozen SSWM encodes the short-horizon consequences of each candidate action under locally plausible teammate and opponent behavior. Each IPPO actor uses online learning to discover how to translate these consequence tokens into an annealed correction of its action preferences. This enables knowledge transfer from offline populations without a centralized critic, communication, or shared online optimizer.

Across 30 matched seeds on Tag, Spread, and Adversary, our in-domain (R+M) variant yields the highest normalized mean performance on every task. Held-out pretraining also yields improved performance over the decentralized baselines on Spread, and remains competitive with the strongest decentralized baselines on Tag and Adversary. Our population comparisons and ablations show that intermediate target-task behavior is more beneficial than using random target-task data alone, and that transient consequence guidance outperforms retaining the model correction throughout training. These results indicate a promising use case for pretrained world models as ephemeral, action-level priors for decentralized online learning instead of persistent controllers or replacements for real experience.

\bibliographystyle{ieeetr}
\bibliography{references}

\section*{Appendix}

\subsection{Software, hardware, and Experimental Setup}
Our implementation used: Python~3.13.7, JAX and jaxlib~0.5.3, Flax~0.10.6, Optax~0.2.6, Distrax~0.1.5, Chex~0.1.90, NumPy~1.26.4, Gymnax~0.0.9. The MPE environments use JaxMARL~0.1.0. GPU support used the JAX CUDA~12 plugin~0.5.3. All our experiments used an NVIDIA L40 GPU.

\begin{reprotable}
\caption{Model architecture and LDWM reference pretraining.}
\label{tab:castle-architecture}
\begin{tabularx}{\linewidth}{@{}Y l@{}}
\toprule
Parameter & Value \\
\midrule
\multicolumn{2}{@{}l}{\textit{Local Dynamics World Model (LDWM)}} \\
Deterministic / hidden dimension & $256$ / $256$ \\
Stochastic variables / categories & $32$ / $32$ \\
Context dimension / uniform mixture & $1280$ / $0.01$ \\
Adam learning rate / epsilon & $3\times10^{-4}$ / $10^{-5}$ \\
Gradient norm limit / KL free nats & $100$ / $0.5$ \\
Obs.\ / reward / continuation weights & $1$ / $1$ / $1$ \\
Dynamics / representation KL weights & $0.8$ / $0.2$ \\
\midrule
\multicolumn{2}{@{}l}{\textit{Semantic-Social World Model (SSWM)}} \\
History length / hidden dimension & $8$ / $128$ \\
Transformer layers / attention heads & $1$ / $4$ \\
Token width / interaction categories & $16$ / $4$ \\
Padded observation dimension & $18$ \\
Entity slots / features per entity & $7$ / $2$ \\
\midrule
\multicolumn{2}{@{}l}{\textit{Actor, Critic, and Guidance}} \\
Attention / feed-forward width & $64$ / $128$ \\
Transformer layers / attention heads & $1$ / $4$ \\
Actor / critic output dimension & $5$ / $1$ \\
Utility hidden width / activation & $64$ / SiLU \\
Utility output per action / gate & $1$ / $0.5\,\sigma(\cdot)$ \\
\bottomrule
\end{tabularx}
\end{reprotable}

\begin{reprotable}
\caption{Online PPO and consequence guidance.}
\label{tab:castle-online-settings}
\begin{tabularx}{\linewidth}{@{}Y l@{}}
\toprule
Parameter & Value \\
\midrule
Parallel environments / rollout steps & $32$ / $128$ \\
Optimizer / epsilon & Adam / $10^{-5}$ \\
Initial learning rate & $3\times10^{-4}$ \\
Learning-rate decay / lower bound & Linear / $10\%$ of initial \\
PPO epochs / minibatches per epoch & $4$ / $4$ \\
Discount / GAE parameter & $0.99$ / $0.95$ \\
Policy clip / gradient norm limit & $0.2$ / $0.5$ \\
Value loss / entropy coefficients & $1.0$ / $0.03$ \\
Guidance coefficient, initial to final & $0.10$ to $0$ \\
Annealing duration / schedule & $244$ updates / linear \\
Token supplied to the actor & Final horizon \\
\bottomrule
\end{tabularx}
\end{reprotable}

World models remain frozen. Guidance gradients do not enter the actor representation. Actor and critic parameters and optimizers are private to each agent.

\begin{reprotable}
\caption{SSWM pretraining, label construction, validation.}
\label{tab:castle-pretraining}
\label{tab:castle-validation}
\label{tab:castle-counterfactual-settings}
\begin{tabularx}{\linewidth}{@{}Y l@{}}
\toprule
Parameter & Value \\
\midrule
\multicolumn{2}{@{}l}{\textit{Optimization and Loss Weights}} \\
Optimizer / learning rate & AdamW / $3\times10^{-4}$ \\
Weight decay / batch size & $10^{-4}$ / $256$ \\
Gradient steps / training and split seed & $10{,}000$ / $0$ \\
Random / decentralized / CTDE sampling & $0.10$ / $0.60$ / $0.30$ \\
Additional final $\Delta$ / $d$ / $c$ weights & $1$ / $1$ / $1$ \\
Horizon-wise $\Delta$ / $d$ / $c$ weights & $0.5$ / $0.5$ / $0.5$ \\
Horizon-wise $R$ / action loss weights & $0.25$ / $0.5$ \\
Token variation / entropy weights & $0.05$ / $0.01$ \\
Token standard-deviation floor & $0.01$ \\
\midrule
\multicolumn{2}{@{}l}{\textit{Label Configuration}} \\
Actions / rollouts ($M$) / horizon ($H$) & $5$ / $4$ / $5$ \\
Counterfactual return discount & $0.95$ \\
Candidate ego action & 1st step \\
Non-ego actions & Sampled behavior policies \\
Subsequent actions, all agents & Sampled behavior policies \\
Rollout aggregation & Mean at each $h$ \\
\midrule
\multicolumn{2}{@{}l}{\textit{Validation}} \\
Training / validation split & $90\%$ / $10\%$ \\
Split procedure & Random permutation \\
Checkpoint evaluated & Final pretraining checkpoint \\
\bottomrule
\end{tabularx}
\end{reprotable}

Sampling is source-stratified with replacement. Weights multiply mean losses; final-horizon terms are additional. The effective entropy weight is $0.01$: $\lambda_{\mathrm{rep}}=0.05$, $\lambda_{\mathrm{ent}}=0.2$. The split is by example, not trajectory.

For each behavior-policy source, we collect 128 trajectories of 25 steps using eight parallel environments.
\begin{reprotable}
\caption{Offline behavior-policy categories.}
\label{tab:offline-population}
\begin{tabularx}{\linewidth}{@{}l X@{}}
\toprule
Source & Category definition \\
\midrule
Random & Uniform action sampling (R). \\
MAPPO & Initial, medium (M), and expert: $0\%$, $30\%$, and $100\%$ of the training budget. \\
Decentralized IPPO & Weak, medium (M), and expert labels assigned from existing performance logs. \\
\bottomrule
\end{tabularx}
\end{reprotable}

\subsection{Task and Evaluation Protocol}
Table~\ref{tab:task-details} presents the number of agents, online training budget, and random and expert return references for each task.

\begin{reprotable}
\caption{Task budgets and raw normalization references.}
\label{tab:task-details}
\begin{tabular}{@{}lcccc@{}}
\toprule
Task & Agents & Timesteps & Random & Expert \\
\midrule
Tag & $4$ & $5$M & $-0.57$ & $36.12$\\
Spread & $3$ & $5$M & $-79.14$ & $-28.46$\\
Adversary & $3$ & $10$M & $-13.26$ & $-1.75$\\
\bottomrule
\end{tabular}
\end{reprotable}
We evaluate each method using 30 matched seeds. For each seed, we average the return logged over the last 200 policy updates to obtain its final raw score. Let $J_{m,e,s}$ denote this score for method $m$, task $e$, and seed $s$. To compare tasks with different reward scales, we normalize each score using the fixed references in Table~\ref{tab:task-details}, following the formula used in D4RL~\cite{fu2020d4rl}:
\begingroup
\small
\begin{equation}
S_{m,e,s} = 100\frac{J_{m,e,s}-J_{\mathrm{random},e}}{J_{\mathrm{expert},e}-J_{\mathrm{random},e}}.
\end{equation}
\endgroup
We report the mean and standard deviation of these scores across seeds. The Total Average is the mean of the three task means. Scores above 100 exceed the corresponding expert reference.
\end{document}